\documentclass[a4paper,11pt]{article}
\usepackage[dvipsnames,table]{xcolor}    

\usepackage{pos}

\usepackage[american]{babel}  
\usepackage[utf8]{inputenc}   
\usepackage[export]{adjustbox}
\usepackage{hyperref}

\graphicspath{{figs}}

\newcommand{\beq}{\begin{equation}}
\newcommand{\eeq}{\end{equation}}

\newcommand{\abs}[1]{\left\vert#1\right\vert}
\newcommand{\kv}[1]

\title{Using AI for Efficient Statistical Inference of Lattice Correlators Across Mass Parameters}

\author*[a]{Octavio Vega}
\author[a]{Andrew Lytle}
\author[a,b]{Jiayu Shen}
\author[a]{Aida X. El-Khadra}

\affiliation[a]{Department of Physics and Illinois Center for Advanced Studies of the Universe\\ University of Illinois Urbana-Champaign, Urbana, IL 61801, USA}

\affiliation[b]{Global Technology Applied Research\\ JPMorganChase, New York, NY 10017, USA}

\emailAdd{octavio5@illinois.edu}
\emailAdd{atlytle@illinois.edu}
\emailAdd{jiayus3@illinois.edu}

\abstract{Lattice QCD is notorious for its computational expense. Modern lattice simulations require large-scale computational resources to handle the large number of Dirac operator inversions used to construct correlation functions. Machine learning (ML) techniques
that can increase, at the analysis level, the information inferred from the correlation functions would therefore be beneficial.
We apply supervised learning to infer two-point lattice correlation functions at different target masses. Our work proposes a new method for separating data into training and bias correction subsets for efficient uncertainty estimation. We also benchmark our ML models against a simple ratio method.}

\FullConference{The 41st International Symposium on Lattice Field Theory (LATTICE2024)\\
 28 July - 3 August 2024\\
Liverpool, UK\\}

\begin{document}
\maketitle

\section{Method}
\begin{figure}
    \centering
    \vspace{-0.04\linewidth}
    \includegraphics[trim = 0 -1cm 0 0, width = 0.58\linewidth]{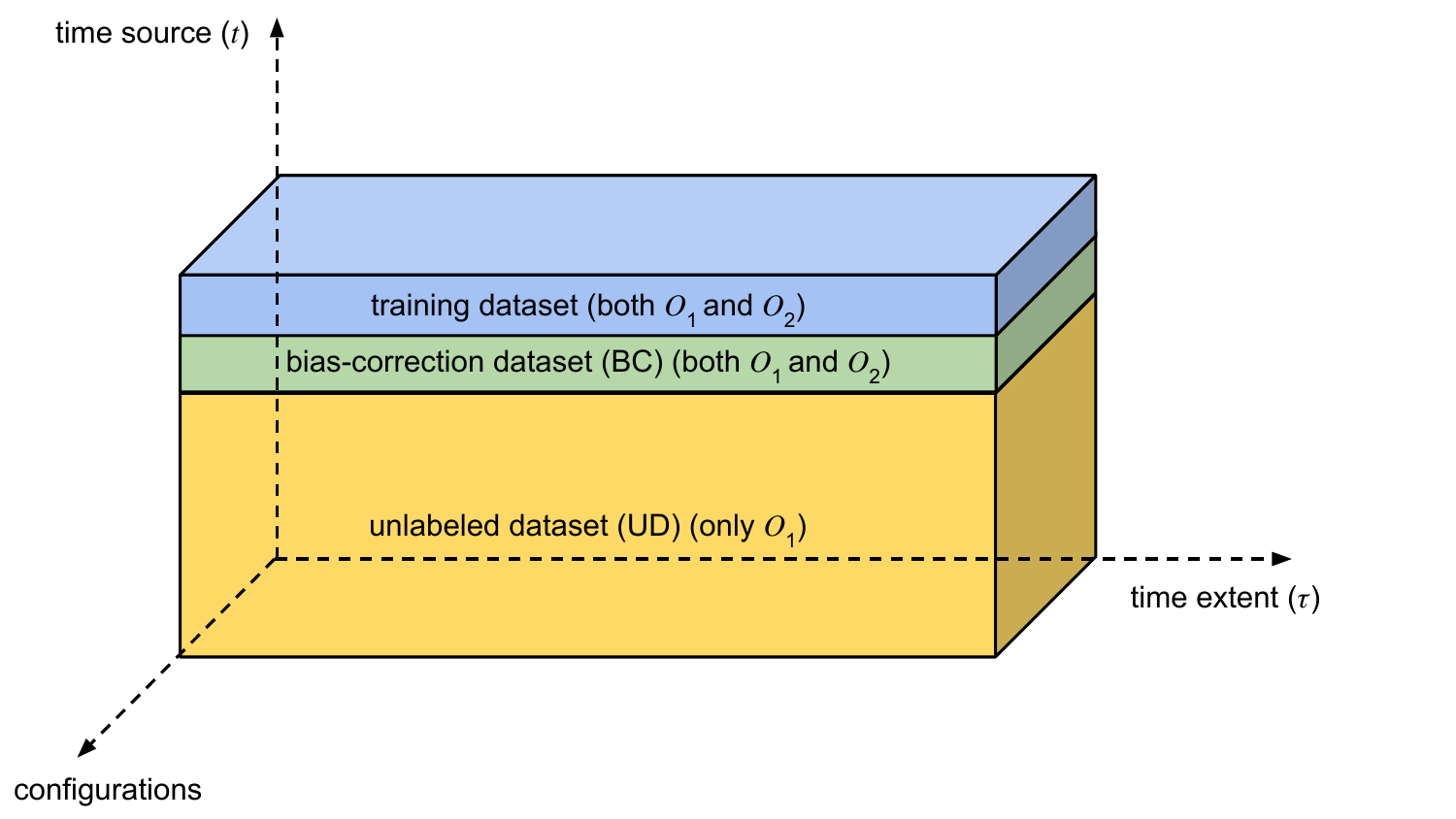}~~
    \hspace{-0.070\linewidth}
    \includegraphics[width = 0.50\linewidth]{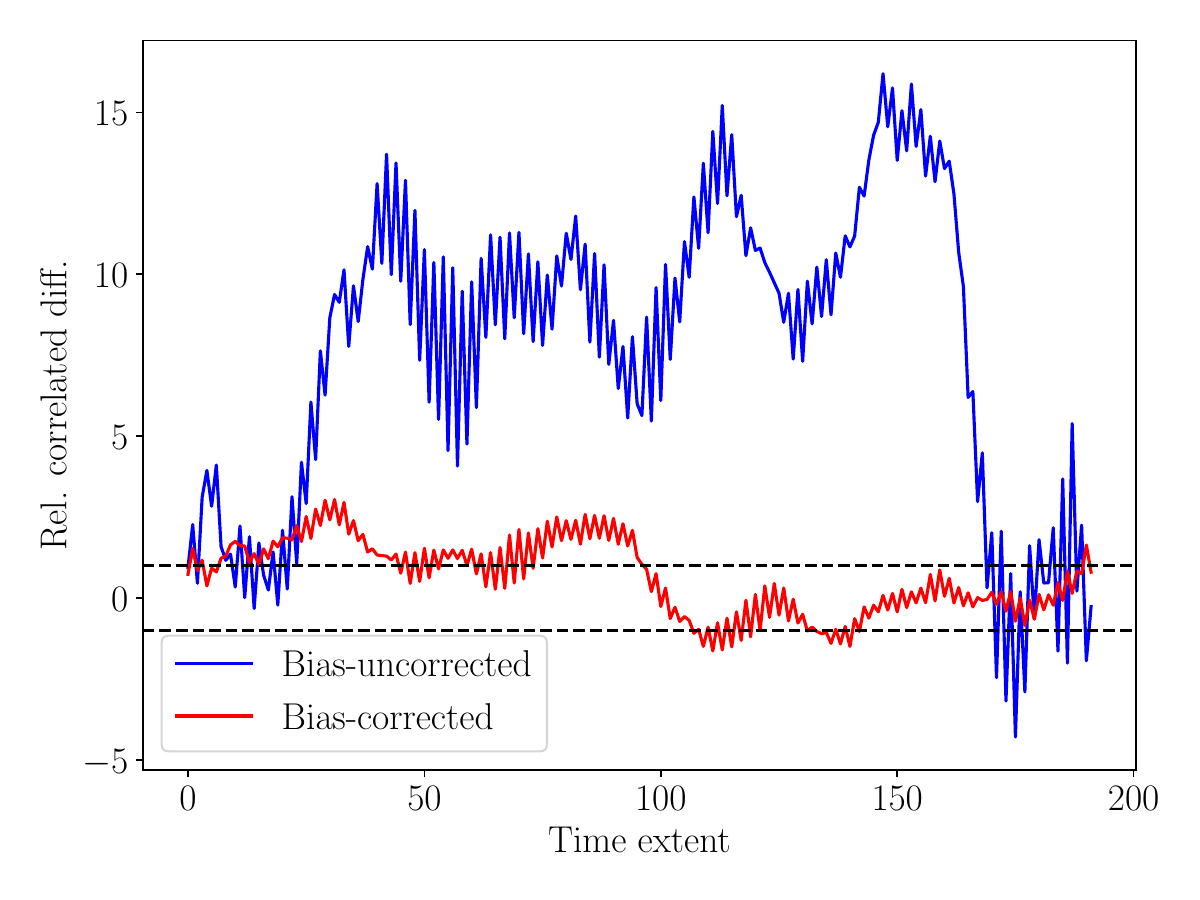}
    \vspace{-0.055\linewidth}
    \caption{(Left) breakdown of the dataset into training, bias-correction, and unlabeled subsets across the source times axis. (Right) impact of bias correction on the relative correlated difference of correlators over the lattice time extent.}
    \label{fig:data-allocation-bias-correction}
\end{figure}

The development of ML methods to analyze lattice correlation functions is an active area of research  (see, for example, \cite{Yoon:2018krb,Kim:2024rpd}).
Adapted from the ML estimator introduced in \cite{Yoon:2018krb}, and inspired by the truncated solver method (TSM)~\cite{Bali:2009hu,Alexandrou:2012zz,Blum:2012uh}, our method uses the source times as indices to separate data into subsets for training, bias-correction, and prediction, thereby preserving the configuration axis to give unbiased configuration-wise predictions. The formula describing our procedure (for time extent $\tau$ and configuration $i$) is
\begin{equation}
    \overline{O_i}(\tau) = \langle O_i(\tau)^\mathrm{pred}\rangle_{\mathrm{UD}} + \langle O_i(\tau) - O_i^
\mathrm{pred}(\tau)\rangle_{\mathrm{BC}}.
\label{eq:oi_bar}
\end{equation}
Here the brackets denote averaging over time sources allocated to the unlabeled (UD) and bias correction (BC) sets, respectively. Writing out the averages explicitly,
\begin{equation}
    \overline{O_i} = \frac{1}{N_\mathrm{UD}} \sum_{j:\,(i, j) \in \mathrm{UD}} O_{i, j}^\mathrm{pred} + \frac{1}{N_\mathrm{BC}} \sum_{k:\,(i, k) \in \mathrm{BC}} (O_{i, k} - O_{i, k}^\mathrm{pred}),
\label{eq:oi_bar_sums}
\end{equation}
where the index pair notation is used to emphasize the data has a two dimensional structure when split into UD, BC, and training sets, and that the configuration axis is preserved. A depiction of the data allocation is shown on the left in Figure~\ref{fig:data-allocation-bias-correction}, and the right plot in Figure~\ref{fig:data-allocation-bias-correction} depicts the effect of bias correction on the correlated difference of predicted correlation functions.

After obtaining the bias-corrected observables $\overline{O_i}$, we can treat them as per-configuration observables in a typical lattice computation, and compute means, covariances, or other statistics. It is also straightforward to use resampling methods on the observables to estimate statistics for more complicated functions of (sets of) observables, for example, to compute correlations in form factors obtained by fitting spectral decompositions. 

The key features of our method are:
\begin{enumerate}
\item The ML predictions are made for each configuration in the ensemble, so that statistical uncertainties can be estimated straightforwardly without the need for intensive bootstrapping or repetitive training.
\item All training source times are ``seen'' by our ML models simultaneously, and network weights are shared across source times for a greater chance to capture their correlations.
\end{enumerate}

\section{Dataset Information}
Our dataset consists of meson two-point functions computed on a single gauge ensemble generated by the MILC collaboration~\cite{MILC:2012znn}, the parameters of which are collected in Table \ref{tab:ens-params}.
\begin{table}[ht]
    \centering
    \begin{tabular}{c c c c c c c c}
    \hline 
    \hline
    \rowcolor{gray!10}
    $ a$ [fm] & $\beta$ & $N_s^3 \times N_t$ & $N_{\text{src}} \times N_{\text{confs}}$ & $m_l/m_s$ &$(am_l)_{\text{sea}}$ & $(am_s)_{\text{sea}}$ & $(am_c)_{\text{sea}}$ \\
    \hline 
    0.042 & 7.00 & $64^3 \times 192$ & $24 \times 1028$ & 1/5 & 0.00316 & 0.0158 & 0.188 \\
    \hline 
    \hline
    \end{tabular}
    \caption{Parameters for the gauge ensemble used in this study. The strange and charm mass parameters have been tuned to close to their physical values. The light mass parameter is taken to be $1/5$ that of the strange, corresponding to a pion mass of $m_\pi \approx$ 308 MeV.}
    \label{tab:ens-params}
\end{table}    
The data was generated as a part of the FNAL-MILC collaboration's $D$- and $B$-meson semileptonic decay program~\cite{FermilabLattice:2022gku}. The valence quark masses used are shown in Table \ref{tab:corr-params}.
\begin{table}[ht]
    \centering
    \begin{tabular}{c c c c c}
    \hline \hline
    \rowcolor{gray!10}
    $(am_l)_{\text{valence}}$ & $(am_s)_{\text{valence}}$ & $(am_h)_{\text{valence}}$&\\
    \hline
    0.00311 & 0.01555 & \{0.164, 0.1827, 0.365, 0.548, 0.731, 0.843 \}&\\
    \hline
    \hline
    \end{tabular}
    \caption{Valence quark mass values used in construction of the two-point correlation functions studied here. The valence light, strange, and charm quark mass parameters are very close to the sea quark mass values listed in Table~\ref{tab:ens-params}. The heavy mass parameters range from 0.9 $m_c$ to 4.2 $m_c$.}
    \label{tab:corr-params}
\end{table}

\section{Correlation functions}
The correlation functions contained in the dataset describe the propagation of a heavy-light (or heavy-strange) meson $H_{(s)}$, created at source time $t_0$ and destroyed at time $t_0 + t$:
\begin{equation}
    C^{2pt}_{H_{(s)}}(t) = \sum_{\mathbf{x}, \mathbf{y}}\langle H_{(s)}(t + t_0, \mathbf{x}) H_{(s)}^{\dagger}(t_0, \mathbf{y}) \rangle ,
\end{equation}
where $H_{(s)}$ is a local staggered interpolating operator with spin-taste structure $\gamma_5 \otimes \gamma_5$.
Its spectral decomposition takes the form:
\begin{equation} \label{spectral_decomposition}
    C^{2pt}_{H_{(s)}}(t) = \sum_{n=0} (-1)^{n (t+1)} \frac{|\langle 0| H_{(s)}^\dagger|n \rangle|^2}{2 E_n} \Bigl(e^{-E_n t} + e^{-E_n (T - t)}\Bigr)
\end{equation}
where the sum is over all energy eigenstates $|n\rangle$ that have the same quantum numbers as the interpolating operator $H_{(s)}$.
\begin{figure}
    \centering
    \includegraphics[width = 0.50\linewidth]{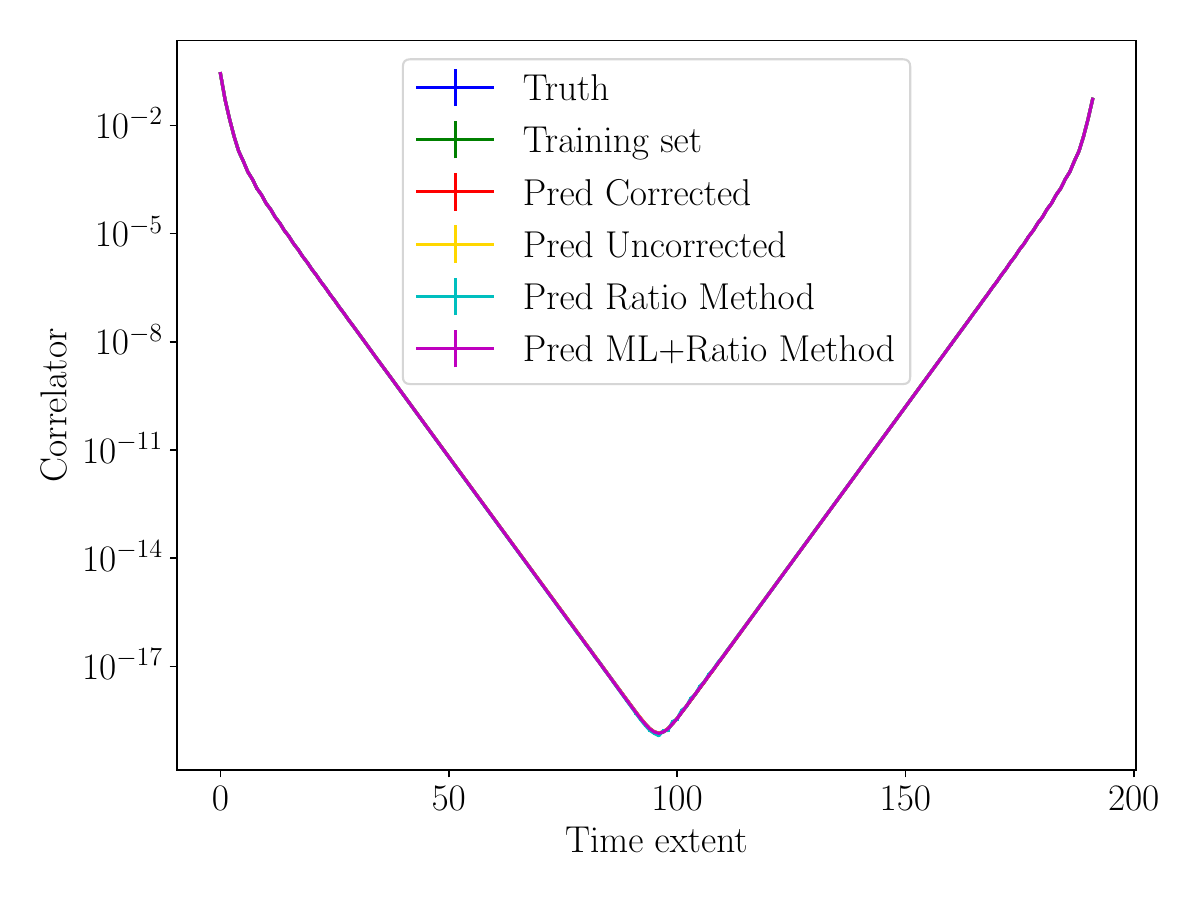}~~
    \includegraphics[width = 0.50\linewidth]{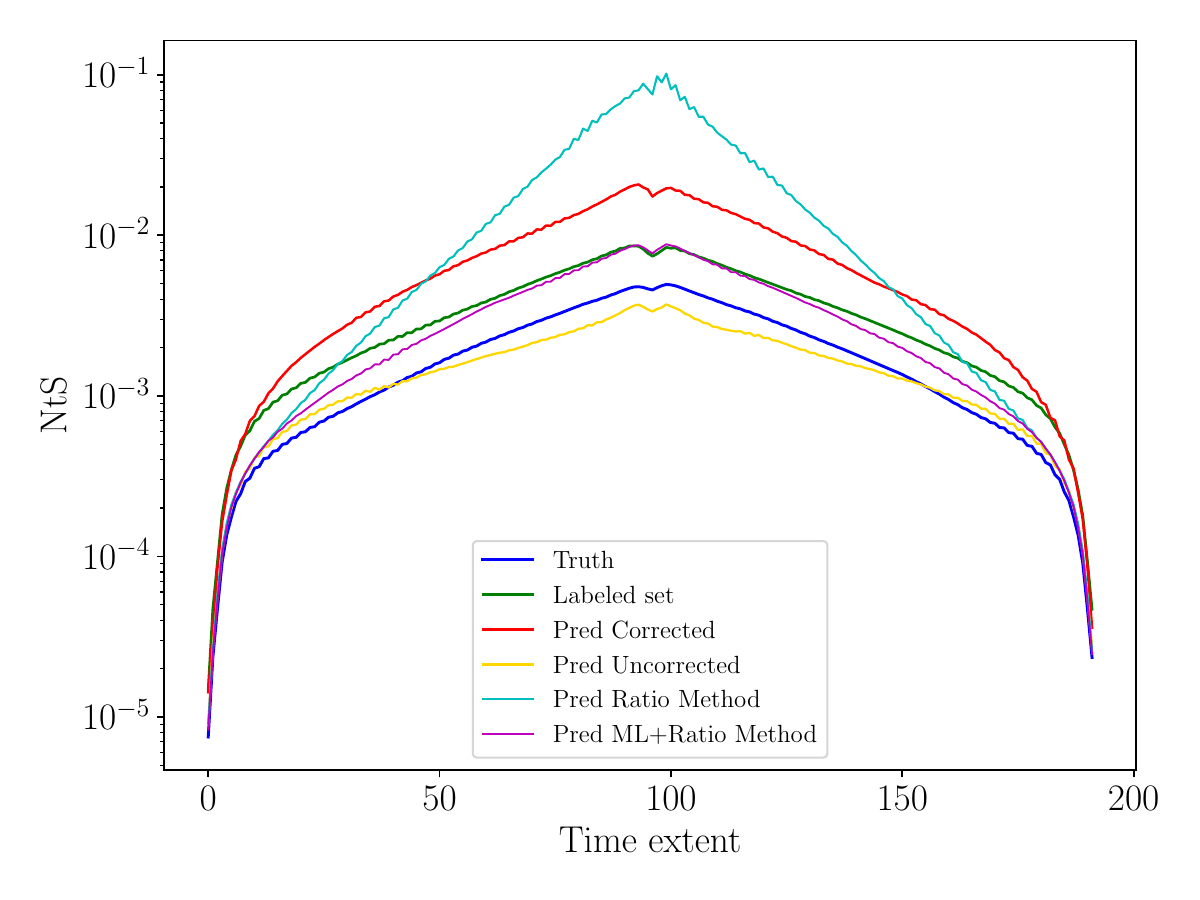}
    \caption{Comparison of predicted correlators (MLP) with truth-level dataset (left) along with their noise-to-signal (NtS) ratios (right) as a function of the Euclidean time extent $\tau$.}
    \label{fig:pred-correlators}
\end{figure}

Spectroscopic information and decay constants can be extracted by fitting the correlator data to the form~\eqref{spectral_decomposition} to determine the few lowest energy eigenvalues, $E_n$, and the overlap amplitudes $|\langle 0| H_{(s)}^\dagger|n \rangle|^2$. Obtaining accurate energies and amplitudes from inferred correlators thus forms a stringent test and benchmark for our inference methods. 

The neural network architectures we use here are the multilayer perceptron (MLP), convolutional neural network (CNN), and the transformer. We also apply a decision tree regressor.
An example of predicted correlators using MLP and their noise-to-signal ratios are plotted in Figure \ref{fig:pred-correlators}.
The training, bias-correction, and unlabeled sets each consist of 1024 configurations selected for each of 1, 5, and 18 (24 total) time source labels. Input and output mass parameters correspond to $m_{i_1} = 0.548$, $m_{i_2} = 0.01555$, $m_{o_1} = 0.164$, and $m_{o_2} = 0.01555$.

\section{Numerical fit results}
We fit our correlator data to the spectral decomposition given in~\eqref{spectral_decomposition}, using the {\tt gvar}~\cite{e}, {\tt lsqfit}~\cite{f}, and {\tt corrfitter}~\cite{g} packages.
We set $t_{\text{min}}=2$ and $n=5$ in Eq.~\eqref{spectral_decomposition}.
The resulting values of amplitudes $a_n$ and energy splitting $dE_n$ for the ground and first excited states from bias-corrected ML predictions compared with results for truth-level data are seen in Table \ref{tab:fit_params}.
{\rowcolors{4}{gray!10}{gray!30}
\begin{table}[h]
    \centering
    \renewcommand\arraystretch{1.3}
    \begin{tabular}{|c|c|c|c|c|c|c|c|}
        \hline
        & \multicolumn{6}{c|}{\textbf{Fit Parameters}}           \\
        \hline
        \textbf{Method} & $a_0$ & $dE_0$ & $a_1$ & $dE_1$ & $\chi^2 / \mathrm{d.o.f.}$ & $Q$ \\
        \hline
        \rowcolor{green!40}
        TRUTH & 0.053791 (43) & 0.398392 (50) &  0.0689 (10) & 0.1838 (21) & 1.07 & 0.30 \\
        MLP & 0.053825 (76) & 0.398386 (78) & 0.0712 (22) & 0.1883 (44) & 0.98 & 0.54 \\
        CNN & 0.053906 (68) & 0.398458 (75) &  0.0740 (18) & 0.1938 (36) & 1.10 & 0.25 \\
        Transformer & 0.053824 (84) & 0.398377 (85) & 0.0713 (23) & 0.1885 (47) & 0.95 & 0.61 \\
        Decision Tree & 0.05382 (10) & 0.39835 (10) & 0.0693 (38) & 0.1854 (72) & 1.09 & 0.27 \\
        \hline
    \end{tabular}
    \caption{Comparisons of fit parameters obtained from the predicted correlators and the truth correlator.}
    \label{tab:fit_params}
\end{table}}

\section{Ratio method as a benchmark}
To benchmark the performance of our machine learning estimators, we make use of ratio estimators, a well-studied method of statistical inference that has been applied before in data-driven lattice studies~\cite{Blum:2023uh}. Ratio estimators empower us to leverage the correlations between observables to estimate a target quantity with improved statistics. 

Concretely, let $O_1$ and $O_2$ be two correlated observables. Then, given $O_1$ computed on all configurations and $O_2$ computed on \textit{some} configurations, we can estimate the value of $O_2$ with a lower uncertainty as follows:
\begin{equation}
    \langle O_2 \rangle_{\text{HP}} = \langle O_1 \rangle_{\text{HP}} \frac{\langle O_2 \rangle_{\text{LP}}}{\langle O_1 \rangle_{\text{LP}}} \,.
\label{eq:ratio_estimator}
\end{equation}
$\langle O \rangle_{\mathrm{HP}} := (1/\abs{\mathcal{S}_{\mathrm{HP}}}) \sum_{i \in \mathcal{S}_{\mathrm{HP}}} O (i)$ is the high-precision (HP) sum over a large subset $\mathcal{S}_{\mathrm{HP}}$, and similarly for $\langle O \rangle_{\mathrm{LP}}$ with $\abs{\mathcal{S}_{\mathrm{LP}}} < \abs{\mathcal{S}_{\mathrm{HP}}}$.  
We also employ a ``boosted'' ratio estimator:
\begin{equation}
    \langle O_2 \rangle_{\text{HP, boosted}} := \langle O_1 \rangle_{\text{HP}}^\alpha \frac{\langle O_2 \rangle_{\text{LP}}}{\langle O_1 \rangle_{\text{LP}}^\alpha} \,,
\label{eq:boosted_ratio_estimator}
\end{equation}
where the constant $\alpha$ exponentiates the high-precision to low precision ratio for $O_1$ and is tailored to minimize the uncertainty of $\langle O_2 \rangle_{\text{HP, boosted}}$. The (boosted) ratio method combined with ML (``(b)RM+ML'') is formulated as
\begin{equation}
\begin{aligned}\label{eq:b(RM) + ML}
   \langle O_2 \rangle_{\text{HP, (b)RM+ML}} := \langle O_1 \rangle_{\text{HP}}^\alpha \frac{\langle O_2^\mathrm{pred} \rangle_{\text{LP}}}{\langle O_1^\mathrm{pred} \rangle_{\text{LP}}^\alpha} \,.
\end{aligned}
\end{equation}
We tabulate a comparison of the different ratio method performances in Table \ref{tab:ratio-method-results}.
{\rowcolors{4}{gray!10}{gray!30}
\begin{table}[h]
    \centering
    \renewcommand\arraystretch{1.3}
    \begin{tabular}{|c|c|c|c|c|c|c|c|}
        \hline
        & \multicolumn{6}{c|}{\textbf{Fit Parameters}}           \\
        \hline
        \textbf{Method} & $a_0$ & $dE_0$ & $a_1$ & $dE_1$ & $\chi^2 / \mathrm{d.o.f.}$ & $Q$ \\
        \hline
        \rowcolor{green!40}
        TRUTH & 0.053791 (43) & 0.398392 (50) &  0.0689 (10) & 0.1838 (21) & 1.07 & 0.30 \\
        RM & 0.05398(12) & 0.39855(15) & 0.0747(22) & 0.1959(47) & 0.91 & 0.73 \\
        bRM & 0.053821(62) & 0.398351(70) & 0.0729(15) & 0.1913(31) & 1.15 & 0.14 \\
        RM + MLP & 0.053811(64) & 0.398353(70) & 0.0723(17) & 0.1903(34) & 1.13 & 0.18 \\
        bRM + MLP & 0.053802(63) & 0.398346(69) & 0.0713(18) & 0.1884(35) & 1.16 & 0.14 \\
        \hline
    \end{tabular}
    \caption{Fit results from combining the ratio method with machine learning, where the model used in this case is the MLP. (b)RM stands for ``boosted ratio method" as defined in Eq. \eqref{eq:boosted_ratio_estimator}, and (b)RM + MLP signifies the combination of the (boosted) ratio estimator with ML as detailed in Eq. \eqref{eq:b(RM) + ML}.}
    \label{tab:ratio-method-results}
\end{table}}

\begin{figure}[ht]
    \centering
    \includegraphics[width=0.95\textwidth]{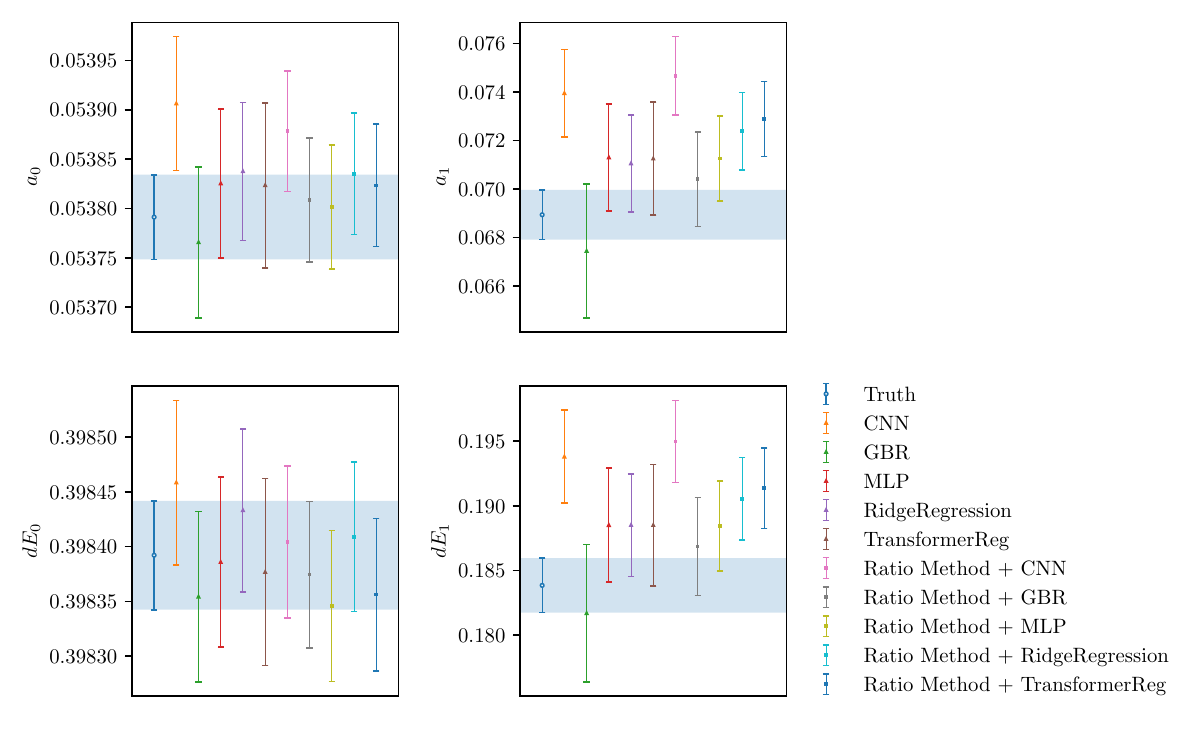}
    \caption{Comparison of fit parameters $a_0$, $a_1$ (top two panels) and $dE_0$, $dE_1$ (bottom two panels) between truth-level data (blue band), bias-corrected ML-predicted data, and ratio method (RM) combined with ML. The ML models shown are the convolutional neural network (CNN), multilayer perceptron (MLP), and gradient-boosted regression tree (GBR).}
    \label{fig:fit_params_comparison}
\end{figure}

The first two amplitude and energy fit parameters are plotted in Figure \ref{fig:fit_params_comparison} and compared for different bias-corrected ML models with ratio method combined with ML as well as with truth data.

\section{Summary}
In this project we develop a new set-up to infer correlation functions with “nearby parameters” and explore a ratio method as well as a range of ML models to predict the correlation functions.  To test the fidelity of our predicted results, we compare them to truth data. Overall, we find that our set-up yields good agreement between the various predictions and the truth.

\acknowledgments
We thank the Fermilab Lattice and MILC collaborations for providing the correlator datasets on which this project is based.
This material is based upon work supported by the U.S. Department of Energy, Office of Science under grant Contract Number DE-SC0015655. A.X.K. and O.V. are grateful to ETH Zürich and the Pauli Center for Theoretical Studies for support and hospitality during summer 2024. O.V. acknowledges support from a University of Illinois Graduate College Fellowship, as well as from a Sloan Scholarship through the Alfred P. Sloan Foundation’s University Center of Exemplary Mentoring, awarded in 2023-2024.

\section*{Disclaimer}
This paper was prepared for informational purposes with contributions from the Global Technology Applied Research center of JPMorgan Chase \& Co. This paper is not a product of the Research Department of JPMorgan Chase \& Co.\ or its affiliates. Neither JPMorgan Chase \& Co.\ nor any of its affiliates makes any explicit or implied representation or warranty and none of them accept any liability in connection with this paper, including, without limitation, with respect to the completeness, accuracy, or reliability of the information contained herein and the potential legal, compliance, tax, or accounting effects thereof. This document is not intended as investment research or investment advice, or as a recommendation, offer, or solicitation for the purchase or sale of any security, financial instrument, financial product or service, or to be used in any way for evaluating the merits of participating in any transaction.

\end{document}